\documentclass[twocolumn,english]{revtex4-2}
\usepackage[T1]{fontenc}
\usepackage[utf8]{inputenc}
\usepackage[english]{babel}

\usepackage{varioref}
\usepackage{float}
\usepackage{units}
\usepackage{amsmath}
\usepackage{amssymb}
\usepackage{graphicx}
\usepackage[unicode=true]{hyperref}
\usepackage{tikz}
\usetikzlibrary{calc}



\begin{document}
\title{Einstein's Equivalence Principle in Nonrelativistic Quantum Mechanics}
\author{S.A.Torres}
\email{satorress@unal.edu.co}

\affiliation{Department of Physics, National University of Colombia}
\date{\today}
\begin{abstract}
	In this work, a precise quantum formulation of Einstein's Equivalence Principle (EEP) is developed within the framework of nonrelativistic quantum mechanics. By employing detailed analyses in both the Schrödinger and Heisenberg pictures, it is demonstrated that an observer in free fall in a uniform gravitational field is equivalent to an inertial observer in the absence of such a field. The transformations of the wave function are reexamined, and the conditions for the equality of inertial and gravitational mass are established---thereby consistently deriving the gravitational redshift observed in neutron interferometry experiments. Additionally, relevant applications of the proposed formalism are explored in confined systems (such as particles in potential boxes and in linear potential barriers), elucidating quantum tunneling phenomena and their dependence on energy levels. The results reaffirm the validity of the Equivalence Principle in the nonrelativistic quantum regime and open new perspectives for integrating quantum mechanics and general relativity. \\

\textbf{Keywords:} Equivalence Principle; Nonrelativistic Quantum Mechanics; Schrödinger Picture; Heisenberg Picture; Gravitational Redshift; Quantum Tunneling; Neutron Interferometry; Linear Potential.
\end{abstract}

\maketitle

\section{Introducción}
Einstein's Equivalence Principle (EEP) postulates that, in the presence of a uniform gravitational field, the motion of an object is indistinguishable from that of another observed from a uniformly accelerating reference frame \citep{Einstein}. This notion, which underpins general relativity, acquires novel dimensions when examined within the framework of nonrelativistic quantum mechanics.

The experimental proposal by Overhauser and Collela \citep{Overhauser-Collela} to verify the EEP via neutron interferometry has enabled the exploration ---on a quantum level--- of the equivalence between free-falling observers and inertial observers in the absence of gravitational fields. Within this context, the work presented in \citep{Nauenberg} provided an analytical solution to the time-dependent Schrödinger equation based on the equality of inertial and gravitational masses. Although this solution has been rederived on several occasions (see Refs. \citep{Staudeman}-\citep{Camacho-Camacho-Guardia}), its analysis has not received the attention it merits in the specialized literature \citep{Eliezer-Leach}.

To address this deficiency, the present work offers a more exhaustive and lucid demonstration of the EEP in quantum mechanics, grounded in precise transformations in both the Schrödinger and Heisenberg pictures. Furthermore, relevant applications of the proposed formalism are examined---highlighting, for instance, the derivation of the gravitational redshift and the analysis of quantum tunneling in confined systems. This approach not only reaffirms the validity of the Equivalence Principle in the nonrelativistic quantum regime but also lays the foundation for future studies aimed at integrating quantum mechanics with general relativity.

\section{EINSTEIN'S EQUIVALENCE PRINCIPLE IN CLASSICAL MECHANICS AND NONRELATIVISTIC QUANTUM MECHANICS}

\subsection{Classical Treatment}
Consider a uniform gravitational field directed along the negative \(z\)-axis. The Newtonian equation of motion for a particle with inertial mass \(m_i\) and gravitational mass \(m_g\) is
\begin{equation}
	m_i\,\frac{d^2 z}{dt^2} = -m_g\,g,\label{eq:Clasica1}
\end{equation}

where \(g\) is the local acceleration due to gravity \citep{Nauenberg}. The equality \(m_i = m_g\) then implies
\begin{equation}
	\frac{d^2 z}{dt^2} = -g.\label{eq:Clasica2}
\end{equation}

Performing a change to a reference frame with coordinate
\begin{equation}
	z' = z + vt + \frac{1}{2}at^2, \quad t' = t,\label{Cambio de Coodenadas}
\end{equation}
where \( v \) is the initial velocity and \( a \) is the constant acceleration of the frame along the negative \(z\)-axis, the equation of motion becomes

\begin{equation}
	\frac{d^2 z'}{dt'^2} = a - \frac{m_g}{m_i}\,g.
\end{equation}
Thus, when
\begin{equation}
	a = \frac{m_g}{m_i}\,g,\label{eq:Clasica3}
\end{equation}
the motion of the particle relative to the accelerated frame (i.e., in a freely falling elevator) is free, with constant velocity determined solely by its initial value.

\subsection{Nonrelativistic Quantum Treatment}

\subsubsection{\textbf{In the Schrödinger Picture}}

\textbf{An observer in free fall within a gravitational field is equivalent to an inertial observer in the absence of gravity: } 

The time-dependent Schrödinger equation for a particle under the action of a constant gravitational potential \(V(z) = m_g\,g\,z\) along the negative \(z\)-axis is:

\begin{equation}
	i\hbar\,\frac{\partial \Psi}{\partial t} = -\frac{\hbar^2}{2m_i}\,\frac{\partial^2 \Psi}{\partial z^2} + m_g\,g\,z\,\Psi.\label{Schrodinger}
\end{equation}
Notably, even when \(m_i = m_g\), the mass appears explicitly in the equation---a marked contrast with the classical case (See Eqs. (\ref{eq:Clasica1})
y (\ref{eq:Clasica2})). However, as will be shown, this mass dependence does not constitute a violation of Einstein's Equivalence Principle.

By changing to the accelerated coordinates
\begin{equation}
	z' = z + vt + \frac{1}{2}at^2,\quad t' = t,
\end{equation}
where \( v \) is the initial velocity and \( a \) is the constant acceleration of the frame along the negative \(z\)-axis, the Schrödinger equation transforms into
\begin{multline}
	i\hbar\left[\frac{\partial\Psi}{\partial t'}+\left(v+at'\right)\frac{\partial\Psi}{\partial z'}\right]=-\frac{\hbar^{2}}{2m_{i}}\frac{\partial^{2}\Psi}{\partial z'^{2}}\\
	+m_{g}g\left(z'-vt'-\frac{1}{2}at'^{2}\right)\Psi\label{Schrodinger en aceleradas}
\end{multline}

We assume a solution of the form
\begin{equation}
	\Psi(z',t') = \Psi'(z',t')\,e^{iS(z',t')},\label{Transformacion}
\end{equation}
with \(\Psi'\) satisfying the free-particle Schrödinger equation in the accelerated frame,

\begin{equation}
	i\hbar\,\frac{\partial \Psi'}{\partial t'} = -\frac{\hbar^2}{2m_i}\,\frac{\partial^2 \Psi'}{\partial z'^2},\label{Schrodinger libre}
\end{equation}

and the function \( S(z', t') \) is a real phase function that ensures the probabilistic interpretation \( |\Psi|^2 = |\Psi'|^2 \). Physically, \( S(z', t') \) accounts for the phase shift due to the gravitational field and the acceleration of the frame. By requiring that the transformation cancels the gravitational potential, one finds that cancellation of the remaining terms demands
\begin{equation}
	a = \frac{m_g}{m_i}\,g,
\end{equation}

aligning with the classical condition for free fall.\\

A proof of this is shown as follows: Substituting the solution (\ref{Transformacion}) into Eq. (\ref{Schrodinger en aceleradas}) yields

\begin{multline}
i\hbar\frac{\partial\Psi'}{\partial t'}=-\frac{\hbar^{2}}{2m_{i}}\frac{\partial^{2}\Psi'}{\partial z'^{2}}-i\hbar\left(\frac{\hbar}{m_{i}}\frac{\partial S}{\partial z'}+\left(v+at'\right)\right)\frac{\partial\Psi'}{\partial z'}\\
+\left\{ \hbar\left[\frac{\partial S}{\partial t'}+\left(v+at'\right)\frac{\partial S}{\partial z'}\right]+\frac{\hbar^{2}}{2m_{i}}\left[\left(\frac{\partial S}{\partial z'}\right)^{2}-i\frac{\partial^{2}S}{\partial z'^{2}}\right]\right.\\
\left.+m_{g}g\left(z'-vt'-\frac{1}{2}at'^{2}\right)\right\} \Psi'\label{Schrodinger Extendida}
\end{multline}

In order for Eqs.  (\ref{Schrodinger}) and (\ref{Schrodinger libre}) to be satisfied, the coefficients of $\Psi'$ and $\nicefrac{\partial\Psi'}{\partial z'}$ in Eq. (\ref{Schrodinger Extendida}) must vanish.

This requirement implies that the function \(S(z',t')\) must be of the form
\begin{equation}
	S(z',t') = -\frac{m_i}{\hbar}(v+at')z' + f(t'), \label{eq:S1}
\end{equation}
where \(f(t')\) is a function solely of \(t'\). Substituting this form into Eq. (\ref{Transformacion}) leads to the relation

\begin{multline}
	\hbar\dot{f}-\frac{1}{2a}\left(m_{i}a+m_{g}g\right)\left(v+at'\right)^{2}+\frac{m_{g}g}{2a}v^{2}\\
	=\left(m_{i}a-m_{g}g\right)z'=\text{constant},\label{eq:16}
\end{multline}

with \(\dot{f} \equiv \frac{df(t')}{dt'}\). Since this equality must hold for all \(z'\in\mathbb{R}\), it follows that \(m_i a - m_g g=0\). That is, the cancellation of the coefficient of \(\Psi'\) requires

\begin{equation}
	a=\frac{m_g\,g}{m_i}, \label{Cuantica1}
\end{equation}
which is exactly the same condition obtained in classical mechanics for uniformly accelerated motion in the primed reference frame (\ref{eq:Clasica3}). Substituting Eq. (\ref{Cuantica1}) into Eq. (\ref{eq:16}) yields the final expression for \(S(z',t')\):

\begin{equation}
	S(z',t') = -\frac{m_i v}{\hbar}\left(z'-\frac{1}{2}vt'\right) - \frac{m_i a t'}{\hbar}\left(z'-vt'-\frac{1}{3}at'^2\right). \label{S}
\end{equation}

Thus, the solution \(\Psi\) of the Schrödinger equation in the accelerated system (with coordinates \((z',t')\)) is

\begin{multline}
\Psi\left(z',t'\right)=\Psi'\left(z',t'\right)\exp\left\{ -\frac{i}{\hbar}\left[m_{i}v\left(z'-\frac{1}{2}vt'\right)\right.\right.\\
\left.\left.+m_{i}at'\left(z'-vt'-\frac{1}{3}at'^{2}\right)\right]\right\} \label{eq:20}
\end{multline}

and the corresponding solution in the stationary coordinate system \((z,t)\), where the gravitational potential is \(V(z)=m_g\,g\,z\) (with \(a=g\) for \(m_i=m_g\)), is

\begin{multline}
\Psi\left(z,t\right)=\Psi'\left(z+vt+\frac{1}{2}at^{2},t\right)\\
\times\exp\left\{ -\frac{i}{\hbar}\left[m_{i}v\left(z+\frac{1}{2}vt\right)+m_{i}at\left(z+\frac{1}{2}vt+\frac{1}{6}at^{2}\right)\right]\right\} \label{eq:21}
\end{multline}

Here, the relation between the gravitational constant \(g\) and the coordinate acceleration relative to it is given by Eq. (\ref{Cuantica1}).

\medskip{}

In summary, it has been demonstrated that there exists a real function \(S(z',t')\), given by Eq. (\ref{S}), such that a wave function \(\Psi'(z',t')\) exists which satisfies the Schrödinger equation for a free particle in the accelerated reference frame \((z',t')\).

The equality of the inertial and gravitational masses, \(m_i=m_g\), thus leads to \emph{Einstein's Equivalence Principle in Nonrelativistic Quantum Mechanics}, with \(a=g\) \citep{Nauenberg}. Substituting Eq. (\ref{eq:21}) for \(\Psi\) into Eq. (\ref{Schrodinger}) shows that \(\Psi\) satisfies the Schrödinger equation for a particle moving in a potential \(V(z)=m_i\,a\,z\). \\

\textbf{An accelerated observer is equivalent to an inertial observer in the presence of a gravitational field: } 

To complete the proof of this equivalence, consider the form of the Schrödinger equation for a free particle in an inertial reference frame relative to an accelerated frame, as defined by the coordinate transformation in Eq. (\ref{Cambio de Coodenadas}). One can show, by a transformation analogous to that in Eq. (\ref{Transformacion}), that in this accelerated frame the Schrödinger equation acquires an additional potential term \(V(z)=m_i\,a\,z\) \citep{Nauenberg}.

\medskip{}

Indeed, let us consider a wave function \(\Psi'\) that satisfies the Schrödinger equation for a free particle in an inertial frame \((z',t')\) relative to an accelerated frame \((z,t)\), as prescribed by the coordinate transformation in Eq. (\ref{Cambio de Coodenadas}):
\begin{equation}
	i\hbar\,\frac{\partial \Psi'}{\partial t'} = -\frac{\hbar^2}{2m_i}\,\frac{\partial^2 \Psi'}{\partial z'^2}.
\end{equation}

Following steps analogous to those employed earlier, one arrives at the conclusion that the Schrödinger equation for the wave function
\begin{equation}
	\Psi(z,t)=\Psi'(z,t)e^{iS(z,t)}
\end{equation}

in the accelerated coordinate system \((z,t)\) takes the form

\begin{equation}
i\hbar\,\frac{\partial \Psi}{\partial t} = -\frac{\hbar^2}{2m_i}\,\frac{\partial^2 \Psi}{\partial z^2} + m_i\,a\,z\,\Psi, \label{eq:Schrodinger_acelerado}
\end{equation}

so that, in the case where \(a=g\), an observer in the accelerated frame will find that \(m_i=m_g\), where \(m_g\) is the gravitational mass.

\medskip{}

All of the foregoing confirms the validity of the \emph{Einstein Equivalence Principle in the Framework of Nonrelativistic Quantum Mechanics:} \\

\textbf{Einstein's Equivalence Principle in Nonrelativistic Quantum Mechanics:} \textit{In the nonrelativistic quantum treatment, a freely falling observer in a gravitational field is equivalent to an inertial observer free of gravitational fields; conversely, an accelerated observer is equivalent to an inertial observer in the presence of a gravitational field.}

\medskip{}

Note that in the case \(a = g = 0\), the Schrödinger equation (\ref{Schrodinger}) reduces to that of a free particle in both coordinate systems \((z,t)\) and \((z',t')\), respectively:
\begin{equation}
	i\hbar\frac{\partial\Psi}{\partial t} = -\frac{\hbar^{2}}{2m_{i}}\frac{\partial^{2}\Psi}{\partial z^{2}}, \label{eq:Schrodinger libre inercial 1}
\end{equation}
\begin{equation}
	i\hbar\frac{\partial\Psi'}{\partial t'} = -\frac{\hbar^{2}}{2m_{i}}\frac{\partial^{2}\Psi'}{\partial z'^{2}}, \label{eq: Schrodinger libre inercial tilde}
\end{equation}

with the solutions in the inertial coordinate systems \((z,t)\) and \((z',t')\) given, respectively, by:
\begin{equation}
	\Psi(z,t)=\Psi'\Bigl(z+vt,t\Bigr)\exp\left\{-\frac{im_{i}v}{\hbar}\left(z+\frac{1}{2}vt\right)\right\}, \label{eq:1.2.25}
\end{equation}
\begin{equation}
	\Psi(z',t')=\Psi'\Bigl(z',t'\Bigr)\exp\left\{-\frac{im_{i}v}{\hbar}\left(z'-\frac{1}{2}vt'\right)\right\}. \label{eq:1.2.24}
\end{equation}

The equivalence of Eqs. (\ref{eq:Schrodinger libre inercial 1}) and (\ref{eq: Schrodinger libre inercial tilde}) demonstrates the invariance of the Schrödinger equation between inertial frames, thereby reflecting the validity of Galileo's Relativity Principle, as expected \citep{Landau-Lifshitz,Viola and R. Onofrio}.

This shows that, \emph{in the nonrelativistic quantum regime, when \(a = g = 0\), Einstein's Equivalence Principle reduces to Galileo's Relativity Principle}.

\medskip{}

\paragraph{\large \textbf{Physical Interpretation of the Transformation (\ref{eq:21})}}
Now consider the plane wave solution for Eq.(\ref{Schrodinger libre}):

\begin{equation}
	\Psi'(z',t') = \Psi'_0\,e^{i\left(k'z' - \omega't'\right)}, \label{Onda plana}
\end{equation}
where \(k'\) and \(\omega'\) are the wave number and its frequency, respectively, for a free particle in an inertial frame (or equivalently, in free fall within a gravitational field). Then, Eq. (\ref{eq:21}) becomes

\begin{multline}
	\Psi(z,t) = \Psi'_0 \exp\Biggl\{ i\Biggl[\left(k' - \frac{m_i v}{\hbar}\right)z - \left(\omega' - v\left[k' - \frac{m_i v}{2\hbar}\right]\right)t \\
	-\frac{m_i a t}{\hbar}z + \frac{a t^2}{2}\left(k' - \frac{m_i v}{\hbar} - \frac{m_i a t}{3\hbar}\right)\Biggr]\Biggr\}, \label{eq:25}
\end{multline}

where the first term of the quantity,

\begin{equation}
	-\frac{m_i a t}{\hbar}z + \frac{a t^2}{2}\left(k' - \frac{m_i v}{\hbar} - \frac{m_i a t}{3\hbar}\right),
\end{equation}

in the argument of the complex exponential in Eq. (\ref{eq:25}) describes a phase contribution associated with the acceleration \(a\) multiplied by time \(t\) and position \(z\). It indicates how the acceleration affects the phase of the wave function along the \(z\)-axis as a function of time. The presence of \(\frac{m_i}{\hbar}\) suggests the influence of the system's inertial mass on the relationship between acceleration and position. \\

The second term represents a quadratic contribution to the phase associated with the acceleration. The presence of \(t^2\) indicates a quadratic dependence on time, which is consistent with the description of a kinematic acceleration term. The detailed expression reveals how the acceleration \(a\) influences the phase as a function of time and other parameters. The appearance of \(\frac{m_i v}{\hbar}\) and \(\frac{m_i a t}{3\hbar}\) suggests the impact of both velocity and acceleration in establishing a quadratic relation between time and phase. \\

Furthermore, the coefficients of \(z\) and \(t\) in the argument of the exponential can be interpreted as follows:

1. \textbf{Term \(\left(k' - \frac{m_i v}{\hbar}\right)z\):} This term describes a phase associated with spatial motion along the \(z\)-axis. The constant \(k'\) is related to the wave number corresponding to spatial position, while \(\frac{m_i v}{\hbar}\) reflects the contribution of momentum to the phase. This term indicates how the spatial coordinate \(z\), the wave number \(k'\), and the momentum \(\left(m_i v/\hbar\right)\) affect the phase of the wave function.

2. \textbf{Term \(\left(\omega' - v\left[k' - \frac{m_i v}{2\hbar}\right]\right)t\):} This term describes a phase associated with the passage of time \(t\). Here, \(\omega'\) is linked to the angular frequency, \(v\) represents the velocity, and \(k'\) again corresponds to the wave number, while \(\frac{m_i v}{2\hbar}\) reflects the momentum contribution to the temporal phase. Together, this term illustrates how the angular frequency \(\omega'\), velocity \(v\), wave number \(k'\), and momentum combine to influence the temporal evolution of the phase of the wave function. \\

\paragraph{\large \textbf{Some Experimental Verifications of the Phase Shift in Equation \eqref{eq:25}:}}

\paragraph{\textbf{Phase Shift:}}

According to Eq. (\ref{eq:25}), the phase shift is given by
\begin{equation}
	\Delta_{Ph}=\left|\frac{m_{i}at}{\hbar}z\right|, \label{Corrimiento de fase}
\end{equation}

\begin{figure}[H]
	\centering
	\includegraphics[width=0.9\linewidth]{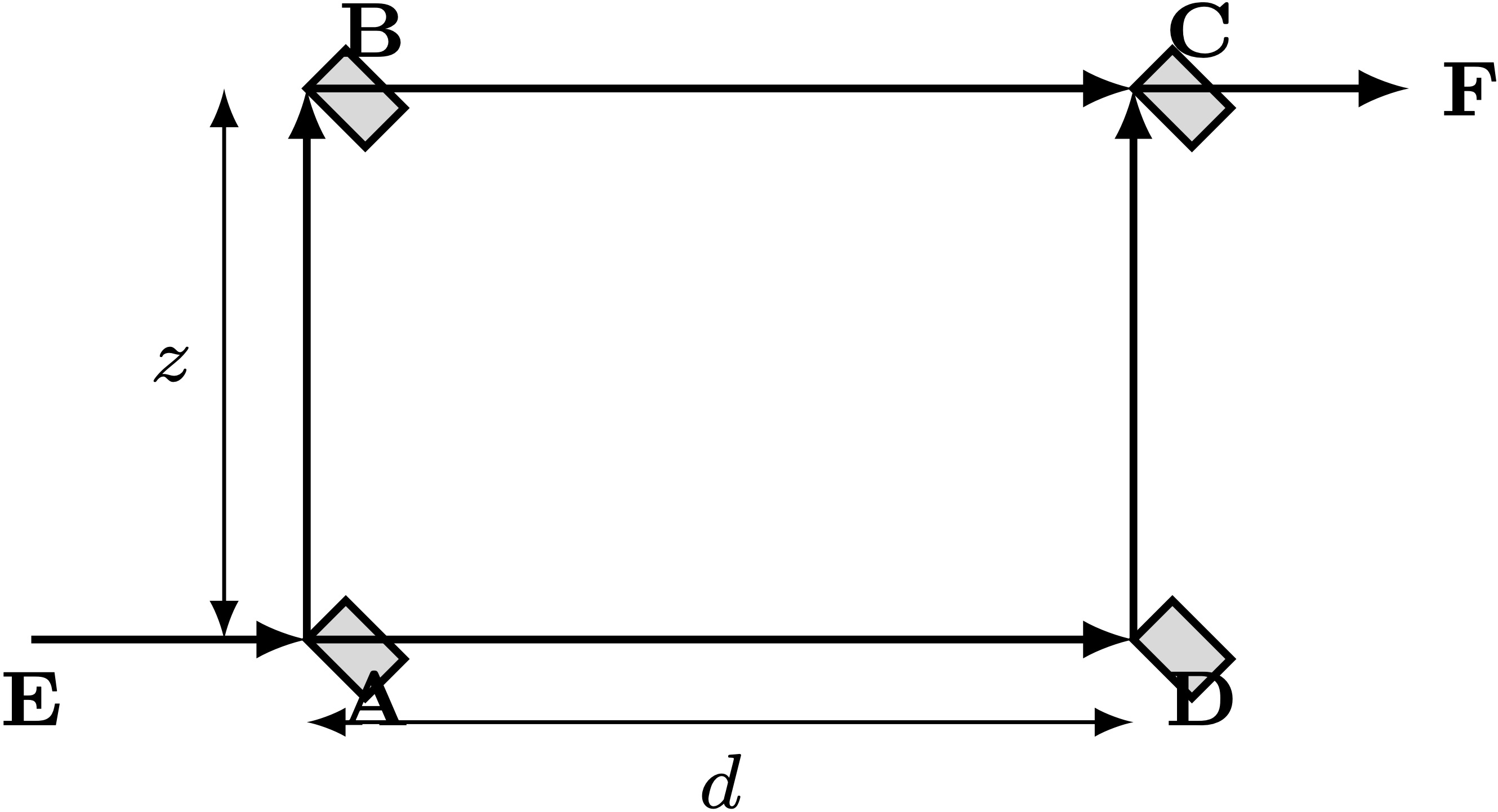}
	\caption{Schematic diagram of a neutron interferometer. The incoming beam (\textbf{E}) is split by mirrors at points \textbf{A}, \textbf{B}, \textbf{C}, and \textbf{D}, and recombined into the outgoing beam (\textbf{F}). The enclosed area is given by \(A = z\,d\).}
	\label{fig:neutroninterferometer}
\end{figure}

In 1974, Overhauser and Collela proposed a neutron interferometry experiment using a grating (see Figure \ref{fig:neutroninterferometer}) to experimentally verify whether neutrons satisfy the predictions of the equivalence principle in quantum mechanics \citep{Overhauser-Collela}.

Taking \(t=\nicefrac{d}{v}\), where \(v=\nicefrac{2\pi\hbar}{m_{i}\lambda}\) is the horizontal velocity of the neutrons and \(\lambda\) their wavelength, Eq. (\ref{Corrimiento de fase}) yields the phase shift
\begin{equation}
	\Delta_{Ph}=\frac{m_{i}^{2}a\lambda zd}{2\pi\hbar^{2}}=\frac{m_{i}^{2}a\lambda A}{2\pi\hbar^{2}}, \label{eq:corrimiento de fase}
\end{equation}

where \(A=zd\) is the area enclosed by the neutron beams. Setting \(m_{i}=m_{g}\) (by Eq. (\ref{Cuantica1})), one obtains \(a=g\), and the phase shift given in Eq. (\ref{eq:corrimiento de fase}) becomes the expression derived by Overhauser and Collela \citep{Overhauser-Collela}. \\

\paragraph{\textbf{Wavelength \(\lambda(t)\) and Frequency \(\omega(t)\):}}

Consider that in Eq. (\ref{Onda plana}) one has a plane wave with linear momentum 
\begin{equation}
p'=\hbar k'.
\end{equation}

Since the particle is free in an inertial frame (or equivalently, in free fall within a gravitational field), one has
\begin{equation}
\hbar\omega'=\frac{{p'}^{2}}{2m_{i}},
\end{equation}

in that reference frame. Then, Eq. (\ref{Onda plana}) can be rewritten as

\begin{equation}
	\Psi'(z',t')=\Psi'_{0}\exp\!\Biggl(\frac{i}{\hbar}\Bigl[p'z'-\frac{{p'}^{2}}{2m_{i}}t'\Bigr]\Biggr),
\end{equation}

and Eq. (\ref{eq:25}) becomes
\begin{multline}
	\Psi(z,t)=\Psi'_{0}\exp\!\Biggl\{\frac{i}{\hbar}\Biggl[\Bigl(p'-m_{i}v\Bigr)z-\Bigl(\frac{{p'}^{2}}{2m_{i}}-v\Bigl[p'-\frac{m_{i}v}{2}\Bigr]\Bigr)t\\
	-m_{i}a t\,z+\frac{a t^{2}}{2}\Bigl(p'-m_{i}v-\frac{m_{i}at}{3}\Bigr)\Biggr]\Biggr\}.
\end{multline}

It can be verified that \(\Psi\) is an eigenstate of the momentum operator 
\(\hat{p}_{z}:=-\frac{i\hbar\partial}{\partial z}\) and of the Hamiltonian operator 
\(\hat{H}:=\frac{i\hbar\partial}{\partial t}\)
\begin{equation}
	-i\hbar\frac{\partial\Psi}{\partial z}=\Bigl[p'-m_{i}\bigl(v+at\bigr)\Bigr]\Psi(z,t)=p(t)\Psi(z,t),\label{eq:Momentum}
\end{equation}
where 
\begin{equation}
	p(t)\equiv p'-m_{i}\bigl(v+at\bigr)
\end{equation}
 
is the momentum eigenvalue for the stationary observer in a gravitational field, when \(m_{i}a=m_{g}g\) (or equivalently, in a linearly accelerated frame with \(a=\frac{m_{g}g}{m_{i}}\)) \citep{Nauenberg}.

\begin{equation}
	i\hbar\frac{\partial\Psi}{\partial t}=\Bigl[\frac{p(t)^{2}}{2m_{i}}+m_{g}gz\Bigr]\Psi(z,t)=E(z,t)\Psi(z,t),\label{eq:Energia}
\end{equation}
where 

\begin{equation}
	E(z,t)\equiv\frac{p(t)^{2}}{2m_{i}}+m_{g}gz
\end{equation}

is the energy eigenvalue for the stationary observer in a gravitational field, which, by relation (\ref{Cuantica1}) with \(m_{i}=m_{g}\), can be rewritten as
\begin{equation}
	E(z,t)=\frac{p(t)^{2}}{2m_{i}}+m_{i}az.\label{eq:Energia2}
\end{equation}

Using the relation

\begin{equation}
	p(t)=\frac{2\pi\hbar}{\lambda(t)},
\end{equation}

Eq. (\ref{eq:Momentum}) shows that

\begin{equation}
	\frac{2\pi\hbar}{\lambda(t)}=\frac{2\pi\hbar}{\lambda'}-m_{i}\bigl(v+at\bigr),
\end{equation}

and hence, the wavelength dilation is given by
\begin{equation}
	\frac{\Delta\lambda}{\lambda(t)}=\frac{m_{i}\bigl(v+at\bigr)}{2\pi\hbar}\lambda'.
\end{equation}

\medskip{}

On the other hand, using the energy relation

\begin{equation}
	E(z,t)=\hbar\omega(z,t),
\end{equation}

Eq. (\ref{eq:Energia2}) shows that, for an arbitrary but fixed time \(t\), one obtains
\begin{equation}
	\hbar\omega(z,t)-\hbar\omega(0,t)=E(z,t)-E(0,t)=m_{i}az.
\end{equation}

This demonstrates that if two particle detectors are placed at positions separated vertically by \(z\), they detect, at an arbitrary time \(t\), a frequency difference 

\begin{equation}
	\Delta\omega\equiv\omega(z,t)-\omega(0,t)
\end{equation}

between the particles arriving at them, which is independent of time and directly proportional to the potential difference \(az\), namely,
\begin{equation}
	\Delta\omega=\frac{m_{i}az}{\hbar}.\label{Redshift1}
\end{equation}

Now, consider the mass-energy relation \(E=mc^{2}\) and the interpretation of light as a quantum of energy with effective mass \(m_{\text{eff}}\). According to the mass-energy relation, a quantum of matter with energy \(E'=\hbar\omega'\) has a mass

\begin{equation}
	m_{\text{eff}}=\frac{E'}{c^{2}}=\frac{\hbar\omega'}{c^{2}}.
\end{equation}

Replacing this result in Eq. (\ref{Redshift1}) yields

\begin{equation}
	\Delta\omega=\frac{\hbar\omega'az}{\hbar c^{2}}=\frac{\omega'az}{c^{2}},
\end{equation}

which leads to the Doppler red/blue shift equation \citep{Carroll}
\begin{equation}
	\frac{\Delta\omega}{\omega'}=\frac{az}{c^{2}}.\label{eq:Redshift-Doppler}
\end{equation}

According to the Equivalence Principle, an \emph{equivalent} situation should occur in the presence of a uniform gravitational field with \(a=g\). Thus, if a photon is emitted from the ground (\(z_{0}=0\)) with frequency \(\omega'\) toward a detector located at a height \(z>0\), it will exhibit a gravitational redshift given by
\begin{equation}
	\frac{\Delta\omega}{\omega'}=\frac{gz}{c^{2}},\label{eq:Gravitational-Redshift}
\end{equation}
where \(gz\) is the gravitational potential difference between the light source and the detector. This is the well-known formula for gravitational redshift \citep{Carroll}.

\medskip{}

Although both the Doppler redshift (\ref{eq:Redshift-Doppler}) and the gravitational redshift (\ref{eq:Gravitational-Redshift}) are associated with changes in the wavelength and frequency of light, they have different underlying causes. On the one hand, gravitational redshift arises from the curvature of spacetime, whereas Doppler redshift is related to the relative motion between the light source and the observer. Both phenomena can occur simultaneously and add together.

\medskip{}

The physical implications of gravitational redshift are fundamental to our understanding of gravity in the context of general relativity. Some implications are:
\begin{itemize}
	\item \textbf{Wavelength Stretching:} When light travels through a gravitational field, its wavelength is stretched. This stretching manifests as a redshift of the electromagnetic spectrum. The magnitude of the gravitational redshift depends on the strength of the gravitational field experienced by the light.
	\item \textbf{Change in Photon Energy:} Gravitational redshift is associated with the loss of gravitational potential energy of photons as they exit a gravitational field. Consequently, the energy of the photons decreases, which manifests as a redshift in their wavelengths.
	\item \textbf{Effect on the Light from Stars and Galaxies:} In astrophysics, gravitational redshift can affect the light emanating from stars and galaxies situated in strong gravitational fields. This effect has been observed in compact stellar systems \citep{Hod}, black holes \citep{Abuter_Roberto}, and galaxy clusters \citep{Kaiser}.
\end{itemize}

\medskip{}

Gravitational redshift has been measured and observed precisely in various astronomical contexts, such as in stars near the supermassive black hole at the center of our galaxy (Sagittarius A{*}) \citep{Hollywood}, or in studies of galaxy clusters and regions with high mass density \citep{Kaiser, Boselli}.

\medskip{}

Note that the frequency difference in Eq. (\ref{Redshift1}) implies the phase shift described in Eq. (\ref{Corrimiento de fase}). That is, the Doppler red/blue shift produces the phase shift in Eq. (\ref{Corrimiento de fase}). According to the Equivalence Principle, a gravitational redshift produces a phase shift:
\begin{equation}
	\Delta_{Ph}=\left|\frac{m_{i}gt}{\hbar}z\right|,\label{eq:Corrimiento_de_fase_gravitacional}
\end{equation}
which is identical in form to the phase shift produced by the Doppler effect.

It is worth mentioning that several innovative experiments have been performed to verify the Equivalence Principle using particles other than neutrons. These studies have primarily explored the behavior of antimatter and cold atoms in gravitational fields, considerably expanding our understanding of how gravity affects matter at a quantum scale. \\

On one hand, the ALPHA experiment at CERN has been fundamental in studying how gravity affects antimatter. By using antihydrogen---which is neutral and thus ideal for such tests because it is not significantly affected by electromagnetic forces---researchers have observed that antimatter, like antihydrogen, responds to gravity in a manner similar to ordinary matter. This is a crucial finding for fundamental physics \cite{Anderson}. \\

On the other hand, high-precision tests employing cold-atom interferometry have been carried out to examine the Equivalence Principle in a quantum environment. These experiments have significantly improved upon the limits of classical tests by using the well-defined properties of cold atoms to measure gravitational effects with great precision \citep{Yuan}.

\medskip{}

\paragraph{\textbf{Applications of Equation (\ref{eq:21})}}
\begin{enumerate}
	\item \textbf{Particle in a One-Dimensional Potential Box in Free Fall.}
	
	Consider a free particle of inertial mass \(m_{i}\) confined within a one-dimensional potential box of height \(L\), which is in free fall in a uniform linear gravitational field with acceleration 
	\begin{equation}
		a=\frac{m_{g}g}{m_{i}}.
	\end{equation}

	The potential energy to which the particle is subjected, relative to the reference frame in which the particle is free, is given by
	\begin{equation}
		V'\bigl(z'\bigr)=
		\begin{cases}
			\infty, & \text{if } z'<0\quad\text{(Region I)}\\[1mm]
			0, & \text{if } 0\leq z'\leq L\quad\text{(Region II)}\\[1mm]
			\infty, & \text{if } z'>L\quad\text{(Region III)}
		\end{cases}
	\end{equation}
	
	The time-dependent Schrödinger equation for Region II is
	\begin{equation}
		i\hbar\frac{\partial\Psi'_{II}}{\partial t'}=-\frac{\hbar^{2}}{2m_{i}}\frac{\partial^{2}\Psi'_{II}}{\partial z'^{2}},\label{Free falling particle in a box}
	\end{equation}
	with the boundary conditions for the confined particle given by
	\begin{equation}
		\Psi'_{II}(0)=0 \quad\text{and}\quad \Psi'_{II}(L)=0.
	\end{equation}
	
	By applying the method of separation of variables, together with the boundary conditions for the potential \(V'(z')\) as given, the solutions \(\Psi'_{n}\) in \((z',t')\) for Eq. (\ref{Free falling particle in a box}) are found to be
	\begin{multline}
		\Psi'_{n}\bigl(z',t'\bigr)=\sqrt{\frac{2}{L}}\sin\!\left(\frac{n\pi z'}{L}\right)\exp\!\left[-\frac{in^{2}\pi^{2}\hbar}{2m_{i}L^{2}}t'\right],\\
		\quad 0\leq z'\leq L.
	\end{multline}
	with allowed momentum and energy levels
	\begin{align}
		p_{n}' & =\hbar k_{n}'=\frac{n\pi\hbar}{L}=\frac{nh}{2L},\\[1mm]
		E_{n}' & =\hbar\omega_{n}'=\frac{{p_{n}'}^{2}}{2m_{i}}=\frac{n^{2}\pi^{2}\hbar^{2}}{2m_{i}L^{2}}=\frac{n^{2}h^{2}}{8m_{i}L^{2}}.
	\end{align}
	
	According to Eq. (\ref{eq:21}), the corresponding solution in the coordinate system \((z,t)\), where the particle is stationary in a gravitational potential \(V(z)=m_{g}gz\) (with \(a=g\) when \(m_{i}=m_{g}\)), is
	\begin{multline}
		\Psi(z,t)=\sqrt{\frac{2}{L}}\sin\!\left[\frac{n\pi}{L}\left(z+vt+\frac{at^{2}}{2}\right)\right]\\
		\times\exp\!\Biggl\{ -\frac{i}{\hbar}\Biggl[m_{i}vz+t\Biggl\{ \frac{\left(\frac{n\pi\hbar}{L}\right)^{2}}{2m_{i}}+\frac{m_{i}v^{2}}{2}\\[1mm]
		\quad\quad\quad\quad\quad\quad\quad + m_{i}a\left(z+\frac{vt}{2}+\frac{at^{2}}{6}\right)\Biggr\}\Biggr] \Biggr\},\label{eq:Particula en caja en caida libre}
	\end{multline}
	for
	\begin{equation}
		-vt-\frac{at^{2}}{2}\leq z\leq L-vt-\frac{at^{2}}{2},
	\end{equation}

	which represents an eigenstate of both the momentum operator 
	\(\hat{p}_{z}:=-\frac{i\hbar\partial}{\partial z}\) and the Hamiltonian 
	\(\hat{H}:=\frac{i\hbar\partial}{\partial t}\), with eigenvalues:
	\begin{equation}
		p_{n}(t)=\frac{nh}{2L}-m_{i}\bigl(v+at\bigr),
	\end{equation}
	\begin{multline}
		E_{n}(z,t)=\frac{p_{n}(t)^{2}}{2m_{i}}+m_{i}az\\
		=\frac{\Bigl[\frac{nh}{2L}-m_{i}\bigl(v+at\bigr)\Bigr]^{2}}{2m_{i}}+m_{i}az,
	\end{multline}
	respectively.
	
	A detailed and complete physical analysis of the wave function \(\Psi(z,t)\) given in Eq. (\ref{eq:Particula en caja en caida libre}) is now presented.
	
	\paragraph*{Analysis of the Sinusoidal Part:}
	
	The sinusoidal part of the wave function is 
	\begin{equation}
		\sqrt{\frac{2}{L}}\sin\!\left[\frac{n\pi}{L}\left(z+vt+\frac{at^{2}}{2}\right)\right].
	\end{equation}
	
	This term describes the spatial probability distribution of finding the particle within the box. Since the particle is in free fall:
	\begin{itemize}
		\item \(z\): Position within the box, relative to an external observer stationary in the gravitational field.
		\item \(vt\): Displacement due to the initial velocity \(v\) of the box.
		\item \(\frac{at^{2}}{2}\): Displacement due to the constant acceleration \(a=g\).
	\end{itemize}
	
	The argument of the sine function reflects the total displacement of the particle within the box, combining both linear and accelerated motion.
	
	The sinusoidal part therefore reflects how the particle moves within the box under the influence of an initial velocity and a constant acceleration. The effective position \(z'=z+vt+\frac{at^{2}}{2}\) represents a combination of uniform rectilinear and accelerated motion.
	
	\paragraph*{Analysis of the Phase Factor:}
	
	The phase factor is given by
	\begin{multline}
		\exp\!\Biggl\{ -\frac{i}{\hbar}\Biggl[m_{i}vz+t\Biggl\{ \frac{\left(\frac{n\pi\hbar}{L}\right)^{2}}{2m_{i}}+\frac{m_{i}v^{2}}{2}\\[1mm]
		\quad\quad\quad\quad\quad\quad\quad+ m_{i}a\Bigl(z+\frac{vt}{2}+\frac{at^{2}}{6}\Bigr)\Biggr\}\Biggr] \Biggr\},
	\end{multline}
	which can be decomposed into the following significant contributions:
	
	\begin{enumerate}
		\item \(m_{i}vz\): This term represents the phase associated with the particle's linear momentum due to the initial velocity \(v\) of the box.
		\item \(\frac{\left(\frac{n\pi\hbar}{L}\right)^{2}}{2m_{i}}t\): This term is the temporal phase corresponding to the kinetic energy of the particle within the box, excluding the contributions from the initial velocity and acceleration.
		\item \(\frac{m_{i}v^{2}}{2}t\): This term represents the kinetic energy of the particle due to the box's initial velocity.
		\item \(m_{i}a\Bigl(z+\frac{vt}{2}+\frac{at^{2}}{6}\Bigr)t\): This term can be further broken down as:
		\begin{itemize}
			\item \(m_{i}az\): Gravitational potential of the particle at position \(z\), relative to an external observer.
			\item \(m_{i}a\frac{vt}{2}\): Correlation between the acceleration and the initial velocity at time \(t\).
			\item \(m_{i}a\frac{at^{2}}{6}\): A higher-order term representing the combined effect of acceleration over time.
		\end{itemize}
	\end{enumerate}
	
	The phase factor captures both the kinetic energy of the particle (which includes contributions from the initial motion and acceleration) and the effect of the gravitational potential on the phase.
	
	From the above analysis, one can conclude that the wave function \(\Psi(z,t)\) completely describes the quantum state of a particle within a one-dimensional potential box in free fall. It reflects both the classical motion (displacement due to velocity and acceleration relative to an external observer) and the quantum properties (energy and phase) of the particle.
	
\item \textbf{Particle Subjected to a Lower One-Dimensional Potential Barrier in a Gravitational Field.}

Next, we present a case where equation (\ref{eq:21}) cannot be directly applied. Consider a stationary particle of inertial mass \(m_{i}\) subjected to a uniform linear gravitational field. The potential energy experienced by the particle is described by the following potential well:
\begin{equation}
	V(z)=
	\begin{cases}
		\infty, & \text{if } z\leq0\quad\text{(Region I)},\\[1mm]
		Fz, & \text{if } z>0\quad\text{(Region II)},
	\end{cases}\label{Potencial_particula_caja_reposo}
\end{equation}
where \(F\equiv m_{g}g\).	

\begin{figure}[H]
	\centering
	\includegraphics[width=0.6\linewidth]{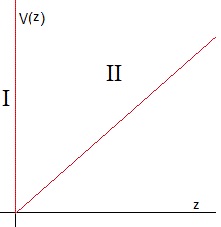}
	\caption{Potential energy for a particle subjected to a lower one-dimensional potential barrier in a gravitational field.}
	\label{fig:potential}
\end{figure}

The time-dependent Schrödinger equation for Region II is
\begin{equation}
	i\hbar\frac{\partial\Psi_{II}}{\partial t}=-\frac{\hbar^{2}}{2m_{i}}\frac{\partial^{2}\Psi_{II}}{\partial z^{2}}+Fz\,\Psi.\label{Schrodinger_en_region_II}
\end{equation}

Since the center of the wave packet is at rest while being subjected to the gravitational field, the particle is not in free fall; hence, equation (\ref{eq:21}) is not applicable.

\textbf{Separation of Variables:} The method of separation of variables is applied by assuming that the solution \(\Psi_{II}(z,t)\) can be written as the product of a function of \(z\) and a function of \(t\):
\begin{equation}
	\Psi_{II}(z,t)=\chi_{II}(z)\,T_{II}(t),
\end{equation}

which leads to the two differential equations
\begin{eqnarray}
	\frac{\hbar^{2}}{2m_{i}}\frac{d^{2}\chi_{II}}{dz^{2}}-\Bigl(Fz-E\Bigr)\chi_{II} & = & 0,\label{Schrodinger_espacial_particula_caja_en_reposo}\\[1mm]
	i\hbar\frac{dT_{II}}{dt} & = & E\,T_{II}.\label{eq:Schrodinger_temporal_particula_caja_en_reposo}
\end{eqnarray}

\textbf{Temporal Solution:} The temporal equation (\ref{eq:Schrodinger_temporal_particula_caja_en_reposo}) has the solution
\begin{equation}
	T_{II}(t)=\exp\!\Bigl(-\frac{iE}{\hbar}t\Bigr).
\end{equation}

\textbf{Spatial Solution:} For the spatial equation (\ref{Schrodinger_espacial_particula_caja_en_reposo}), we perform the change of variable
\begin{equation}
	\xi=\Bigl(\frac{2m_{i}}{\hbar^{2}F^{2}}\Bigr)^{\!\frac{1}{3}}\Bigl(Fz-E\Bigr)
	=\alpha\Bigl(z-\frac{E}{F}\Bigr)
	=\widetilde{z}-\widetilde{E},\label{eq:xi}
\end{equation}
where \(\alpha\equiv\Bigl(\frac{2m_{i}F}{\hbar^{2}}\Bigr)^{\!\frac{1}{3}}\) and the dimensionless quantities are defined as \(\widetilde{z}\equiv\alpha z\) and \(\widetilde{E}\equiv\frac{\alpha E}{F}\). This transformation yields the Airy equation (see pages 446--448 in \citep{Abramowitz M})

\begin{equation}
	\frac{d^{2}\chi_{\widetilde{II}}}{d\xi^{2}}-\xi\,\chi_{\widetilde{II}}=0,\label{eq:Airy}
\end{equation}
in the region \(\widetilde{II}\), defined by \(\xi\geq-\frac{\alpha E}{F}\). The general solution is a linear combination of the Airy functions \(\mathrm{Ai}(\xi)\) and \(\mathrm{Bi}(\xi)\) \citep{Spinel}:
\begin{equation}
	\chi_{II}(z)=A\,\mathrm{Ai}\Bigl(\alpha\Bigl(z-\frac{E}{F}\Bigr)\Bigr)+B\,\mathrm{Bi}\Bigl(\alpha\Bigl(z-\frac{E}{F}\Bigr)\Bigr).
\end{equation}

\textbf{Boundary Conditions:} Since \(\chi_{II}(z)\) must remain finite as \(z\to\infty\) and vanish at \(z=0\), the solution must satisfy
\begin{equation}
	\chi_{I}(0)=\chi_{II}(0)=\chi(0)=0.
\end{equation}

Furthermore, the function \(\mathrm{Bi}(\xi)\) diverges as \(\xi\to\infty\); therefore, the coefficient \(B\) must be zero in order to satisfy the normalization condition. Consequently, the general solution reduces to
\begin{equation}
	\chi_{II}(z)=A\,\mathrm{Ai}\!\Bigl(\alpha\Bigl(z-\frac{E}{F}\Bigr)\Bigr)=A\,\mathrm{Ai}(\widetilde{z}-\widetilde{E}).
\end{equation}

The condition \(\chi(0)=0\) imposes a constraint on the values of \(E\). The allowed values of \(E\) are determined by the zeros of the Airy function \(\mathrm{Ai}(\xi)\):
\begin{equation}
	\mathrm{Ai}\!\Bigl(\alpha\Bigl(-\frac{E}{F}\Bigr)\Bigr)=0,\label{eq:Sol_Particular_Eq_Airy}
\end{equation}

\textbf{Quantized Energy:} From Eq. (\ref{eq:Sol_Particular_Eq_Airy}), we obtain \(\mathrm{Ai}(-\widetilde{E})=0\). Hence, \(-\widetilde{E}\) is a zero of the Airy function \(\mathrm{Ai}(\xi)\). Let
\begin{equation}
	-\widetilde{E}_{n}=-\frac{\alpha E_{n}}{F},\quad n=1,2,3,\dots
\end{equation}

be the zeros of the Airy function. Then, the energy eigenvalues \(E_{n}\) are given by
\begin{equation}
	E_{n}=\widetilde{E}_{n}\left(\frac{\hbar^{2}F^{2}}{2m_{i}}\right)^{\frac{1}{3}}.
\end{equation}

The first energy levels are:
\begin{center}
	\begin{tabular}{|c|c|c|}
		\hline 
		\(n\) & \(\widetilde{E}_{n}\) & \(E_{n}\) \tabularnewline
		\hline 
		\hline 
		\(1\) & \(2.3381\) & \(2.3381\cdot\Bigl(\frac{\hbar^{2}F^{2}}{2m_{i}}\Bigr)^{\frac{1}{3}}\) \tabularnewline
		\hline 
		\(2\) & \(4.0879\) & \(4.0879\cdot\Bigl(\frac{\hbar^{2}F^{2}}{2m_{i}}\Bigr)^{\frac{1}{3}}\) \tabularnewline
		\hline 
		\(3\) & \(5.5206\) & \(5.5206\cdot\Bigl(\frac{\hbar^{2}F^{2}}{2m_{i}}\Bigr)^{\frac{1}{3}}\) \tabularnewline
		\hline 
		\(4\) & \(6.7867\) & \(6.7867\cdot\Bigl(\frac{\hbar^{2}F^{2}}{2m_{i}}\Bigr)^{\frac{1}{3}}\) \tabularnewline
		\hline 
		\(5\) & \(7.9441\) & \(7.9441\cdot\Bigl(\frac{\hbar^{2}F^{2}}{2m_{i}}\Bigr)^{\frac{1}{3}}\) \tabularnewline
		\hline 
		\(6\) & \(9.0227\) & \(9.0227\cdot\Bigl(\frac{\hbar^{2}F^{2}}{2m_{i}}\Bigr)^{\frac{1}{3}}\) \tabularnewline
		\hline 
	\end{tabular}
\end{center}

\begin{table}[H]
	\caption{Energy levels for a particle subjected to a linear potential.\label{tab:Niveles-de-energ=0000EDa}}
\end{table}

\textbf{Normalized Wave Function:} The normalized spatial wave function in Region II is given by
\begin{equation}
	\chi_{n}(z)=A_{n}\,\mathrm{Ai}\!\Bigl(\alpha\Bigl(z-\frac{E_{n}}{F}\Bigr)\Bigr),
\end{equation}

where \(A_{n}\) is a normalization constant such that
\begin{equation}
	\int_{-\infty}^{\infty}\chi^{2}(z)\,dz=\int_{0}^{\infty}\chi_{II}^{2}(z)\,dz=1.
\end{equation}

The temporal wave function is
\begin{equation}
	T_{n}(t)=\exp\!\Bigl(-\frac{iE_{n}}{\hbar}t\Bigr).
\end{equation}

Combining these solutions, the general time-dependent wave function is
\begin{equation}
	\Psi_{n}(z,t)=A_{n}\,\mathrm{Ai}\!\Bigl(\alpha\Bigl(z-\frac{E_{n}}{F}\Bigr)\Bigr)\exp\!\Bigl(-\frac{iE_{n}}{\hbar}t\Bigr).
\end{equation}

\textbf{Graphical Analysis of the Dimensionless Energy Levels:} 

Below, we graphically represent the first few dimensionless energy levels, 
\begin{equation}
	\widetilde{E}_{n}=\frac{\alpha E_{n}}{F},
\end{equation}

and the corresponding wave functions 
\begin{equation}
	\chi_{n}\bigl(\widetilde{z}\bigr)=A_{n}\,\mathrm{Ai}\Bigl(\widetilde{z}-\widetilde{E}_{n}\Bigr),
\end{equation}
in Region \(\widetilde{II}\) for the dimensionless potential well
\begin{equation}
	\widetilde{V}\bigl(\widetilde{z}\bigr)=
	\begin{cases}
		\infty, & \text{if } \widetilde{z}\leq0\quad\text{(Region \(\widetilde{I}\))},\\[1mm]
		\widetilde{z}, & \text{if } \widetilde{z}>0\quad\text{(Region \(\widetilde{II}\))}.
	\end{cases}\label{Pozo_potencial_adimensional}
\end{equation}
The potential well \(\widetilde{V}\bigl(\widetilde{z}\bigr)\) is depicted by thick red lines \citep{Funcion de Airy}.

\begin{figure}[H]
	\centering
	\includegraphics[width=1.1\linewidth]{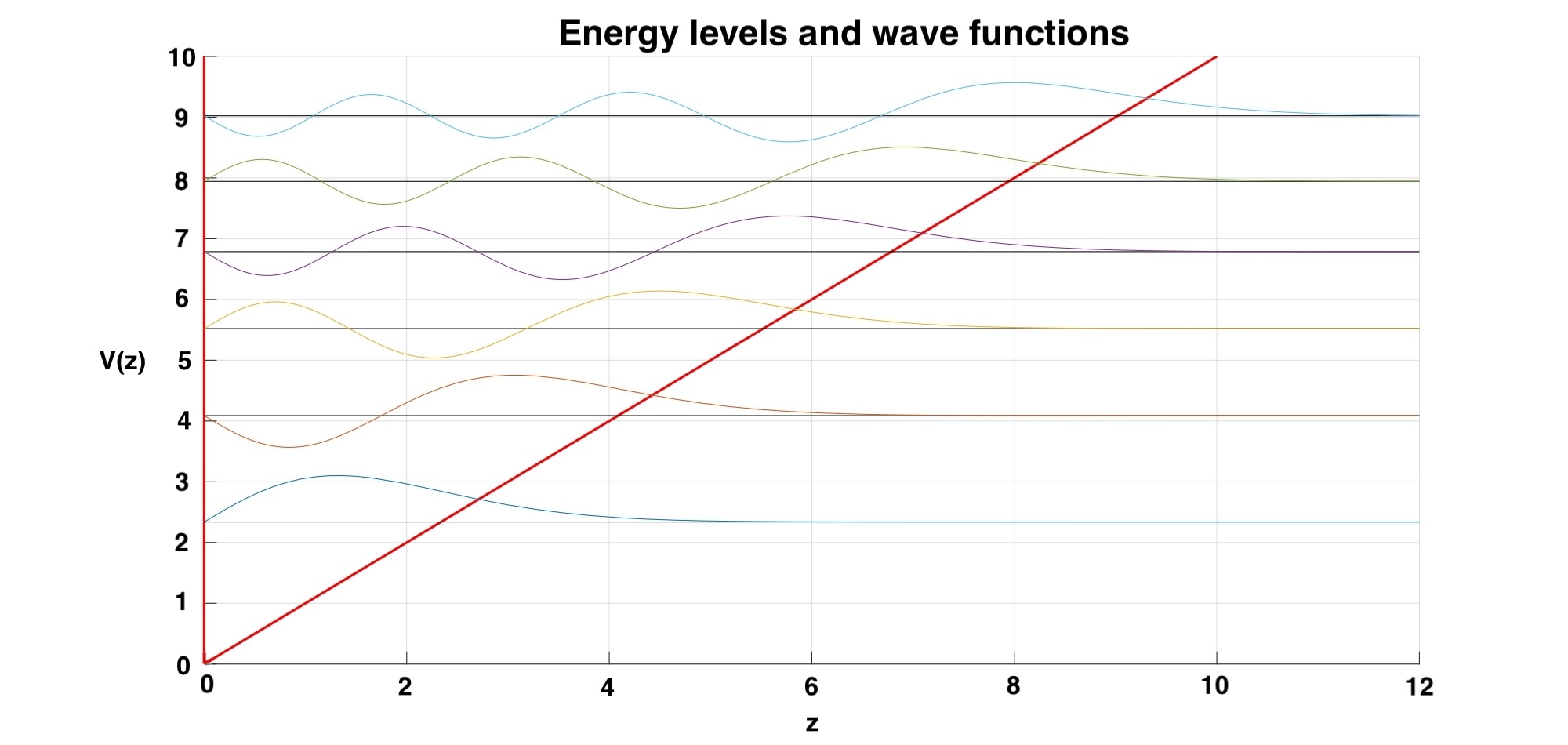}
	\caption{Dimensionless energy levels \(\widetilde{E}_{n}\) and wave functions \(\chi_{n}\bigl(\widetilde{z}\bigr)\).}
	\label{fig:airyfunctions}
\end{figure}

Next, we analyze the probability of finding the particle outside the potential well \(\widetilde{V}\bigl(\widetilde{z}\bigr)\) when it is in the energy level \(\widetilde{E}_{n}\).

\begin{figure}[H]
	\centering
	\includegraphics[width=1.1\linewidth]{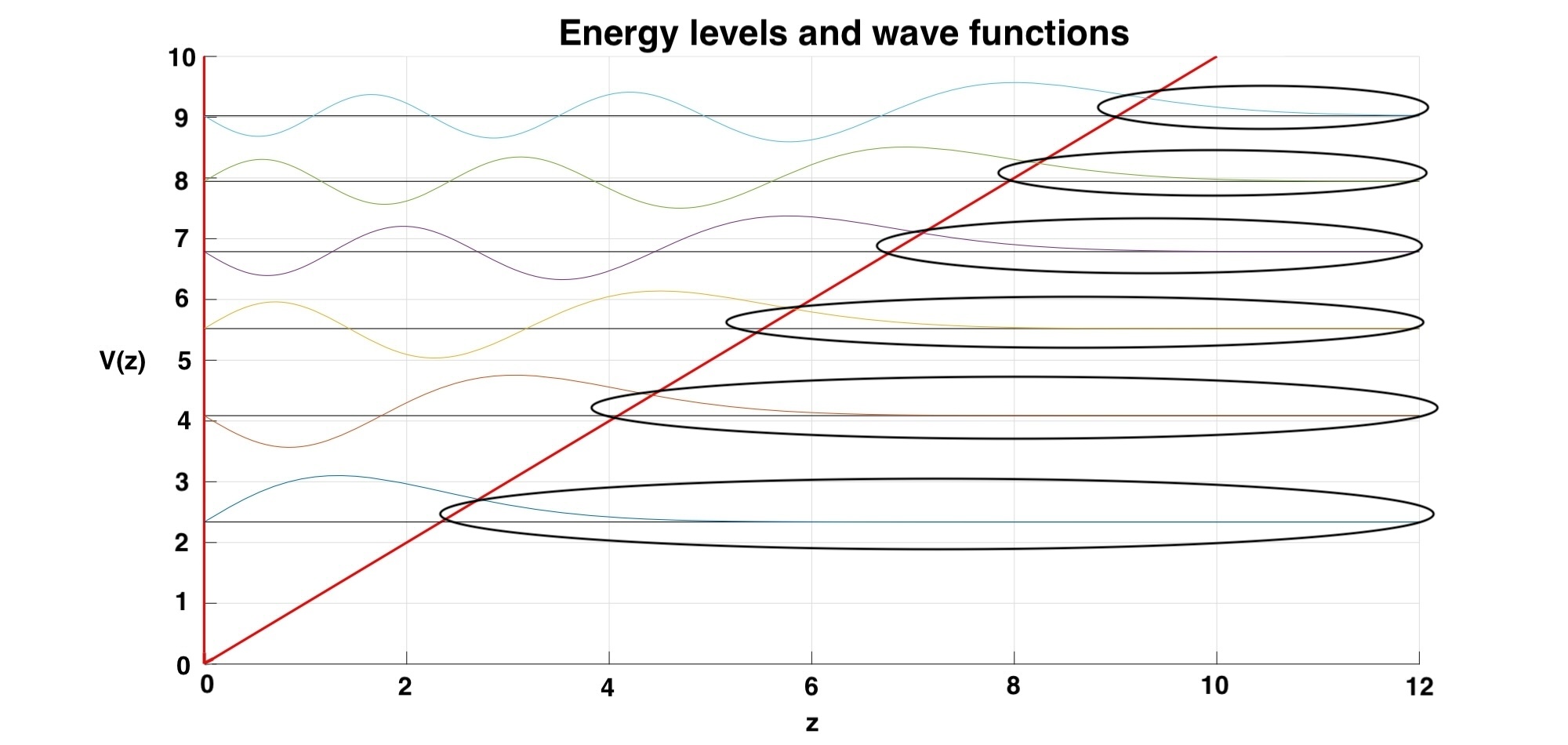}
	\caption{Region where the particle may be found outside the gravitational potential well.}
	\label{fig:airyfunctions2}
\end{figure}

This translates into calculating the probability that the particle is found beyond a certain value of \(\widetilde{z}\), typically where the potential equals the particle's energy, i.e. when \(\widetilde{z}=\widetilde{E}_{n}\).

The probability of finding the particle at \(\widetilde{z}>\widetilde{E}_{n}\) is calculated by integrating the square of the wave function:
\begin{align}
	P\Bigl(\widetilde{z}>\widetilde{E}_{n}\Bigr) & =\int_{\widetilde{E}_{n}}^{\infty}\Bigl|\chi_{n}\bigl(\widetilde{z}\bigr)\Bigr|^{2}d\widetilde{z}\nonumber \\
	& =A_{n}^{2}\int_{\widetilde{E}_{n}}^{\infty}\mathrm{Ai}^{2}\Bigl(\widetilde{z}-\widetilde{E}_{n}\Bigr)d\widetilde{z},\label{eq:probability_1}
\end{align}
and, taking into account the change of variable in Eq. (\ref{eq:xi}), the probability (\ref{eq:probability_1}) can be written as 
\begin{equation}
	P\Bigl(\widetilde{z}>\widetilde{E}_{n}\Bigr)=A_{n}^{2}\int_{0}^{\infty}\mathrm{Ai}^{2}(\xi)d\xi.
\end{equation}

The normalization coefficients \(A_{n}\) are determined by the normalization condition
\begin{align*}
	1 & =\int_{0}^{\infty}\chi_{II}^{2}(z)\,dz\\[1mm]
	& =A_{n}^{2}\int_{0}^{\infty}\mathrm{Ai}^{2}\Bigl(\widetilde{z}-\widetilde{E}_{n}\Bigr)d\widetilde{z}\\[1mm]
	& =A_{n}^{2}\int_{-\widetilde{E}_{n}}^{\infty}\mathrm{Ai}^{2}(\xi)d\xi.\label{normalization_conditions}
\end{align*}
Thus,
\begin{equation}
	A_{n}^{2}=\frac{1}{\displaystyle \int_{-\widetilde{E}_{n}}^{\infty}\mathrm{Ai}^{2}(\xi)d\xi},
\end{equation}

and the probability (\ref{eq:probability_1}) becomes
\begin{equation}
	P\Bigl(\widetilde{z}>\widetilde{E}_{n}\Bigr)=\frac{\displaystyle \int_{0}^{\infty}\mathrm{Ai}^{2}(\xi)d\xi}{\displaystyle \int_{-\widetilde{E}_{n}}^{\infty}\mathrm{Ai}^{2}(\xi)d\xi}.
\end{equation}

For the values given in Table \ref{tab:Niveles-de-energ=0000EDa}, we obtain the following probabilities:
\begin{center}
	\begin{tabular}{|c|c|}
		\hline 
		\(n\) & \(P\Bigl(\widetilde{z}>\widetilde{E}_{n}\Bigr)\) \tabularnewline
		\hline 
		\hline 
		\(1\) & \(13.62\%\) \tabularnewline
		\hline 
		\(2\) & \(10.39\%\) \tabularnewline
		\hline 
		\(3\) & \(8.95\%\) \tabularnewline
		\hline 
		\(4\) & \(8.07\%\) \tabularnewline
		\hline 
		\(5\) & \(7.46\%\) \tabularnewline
		\hline 
		\(6\) & \(7.01\%\) \tabularnewline
		\hline 
		\(7\) & \(6.64\%\) \tabularnewline
		\hline 
		\(8\) & \(6.34\%\) \tabularnewline
		\hline 
		\(9\) & \(6.09\%\) \tabularnewline
		\hline 
		\(10\) & \(5.88\%\) \tabularnewline
		\hline 
	\end{tabular}
\end{center}

These results reflect the probabilities of finding a particle outside the linear potential well for the first ten quantized energy levels. The percentages exhibit a decreasing trend with increasing quantum number \(n\) and are independent of the strength of the uniform gravitational field. We now analyze these results from a physical standpoint to understand what they reveal about the system and the behavior of the particle.

\textbf{Interpretation of the Results}
\begin{itemize}
	\item \textbf{Decay of the Probability with \(n\):} As the quantum number \(n\) increases, the energy levels of the particle also increase. Each higher energy level allows the particle to possess greater energy and, consequently, a larger spatial extent within the potential. However, the probability of finding the particle outside the well, \(P\Bigl(\widetilde{z}>\widetilde{E}_{n}\Bigr)\), decreases. This occurs because, although the particle has more energy, the extension of the wave function into regions where \(\widetilde{z}>\widetilde{E}_{n}\) is comparatively smaller.
	
	\item \textbf{Quantum Tunneling:} The probabilities indicate that there is a nonzero chance for the particle to be present beyond the classical limit set by its dimensionless potential energy \(\widetilde{V}(\widetilde{z})=\widetilde{z}\). This is a manifestation of quantum tunneling, wherein the particle's wave function extends into regions of potential that would be classically forbidden based on its total energy.
	
	\item \textbf{Probabilities and Wave Function Amplitude:} The decrease in probability with increasing energy levels can also be explained by the form of the wave function. At higher energy levels, the wave function tends to exhibit more nodes (points at which the wave function is zero), which implies greater variations in its amplitude. These oscillations may contribute to a more even distribution of probability throughout the well, thereby reducing the amplitude in the "tails" of the wave function that extend into regions where \(\widetilde{z}>\widetilde{E}_{n}\).
\end{itemize}

Furthermore, the relationship between quantum tunneling in a linear potential well and Hawking radiation is an intriguing but indirect connection. Both phenomena involve quantum mechanical principles and the emergence of nonclassical effects, although they occur in very different physical contexts.

On one hand, the quantum tunneling effect discussed in the context of the linear potential well implies that a quantum particle can penetrate a potential barrier even if its classical energy would be insufficient to overcome it. This effect is commonly observed in microscopic structures and is fundamental to devices such as tunnel diodes and field-effect transistors. In the case of a particle in a potential well, quantum tunneling permits the particle to exist in regions where, classically, it would not be allowed based on its energy.

Hawking radiation, on the other hand, is a theoretical phenomenon predicted for black holes. This radiation implies that black holes are not completely black but can emit particles due to quantum effects near the event horizon. The most widely accepted explanation is that particle-antiparticle pairs are continuously created in the vacuum near the event horizon. Under certain conditions, one of these particles may fall into the black hole while the other escapes, resulting in a net loss of mass for the black hole and the emission of detectable radiation.

The conceptual connection between quantum tunneling in a potential well and Hawking radiation lies in the use of quantum theory to describe phenomena that are fundamentally nonclassical:
\begin{itemize}
	\item In both quantum tunneling and Hawking radiation, quantum effects at boundaries or under extreme conditions (such as potential barriers or event horizons) are crucial.
	\item Both phenomena rely on the intrinsic probabilistic nature of quantum mechanics, where events are described by probability amplitudes rather than deterministic certainties.
	\item Both effects are manifestations of the uncertain nature of the quantum world, in which particles can "appear" or "disappear" in unexpected locations.
\end{itemize}

\end{enumerate}

\subsubsection{\textbf{In the Heisenberg Picture}}

\textbf{A free-falling observer in a gravitational field is equivalent to an inertial observer in the absence of a gravitational field:} 

Using the Hamiltonian operator
\begin{equation}
	\hat{H}=\frac{1}{2m_{i}}\hat{p}^{2}+m_{g}g\,\hat{z},\label{eq:Operador_Hamiltoniano}
\end{equation}
in the Heisenberg picture, the Heisenberg equations of motion are
\begin{align}
	\frac{d}{dt}\hat{z}(t) & = -\frac{i}{\hbar}\Bigl[\hat{z}(t),\hat{H}\Bigr] = \frac{1}{m_{i}}\hat{p}(t),\label{eq:Heissenberg_posicion}\\[1mm]
	\frac{d}{dt}\hat{p}(t) & = -\frac{i}{\hbar}\Bigl[\hat{p}(t),\hat{H}\Bigr] = -m_{g}g\,\hat{1},\label{eq:Heissenberg_Momentum}
\end{align}
where the canonical commutation relation
\begin{equation}
	\bigl[\hat{z}(t),\hat{p}(t)\bigr]=i\hbar\,\hat{1}
\end{equation}

has been assumed and \(\hat{1}\) is the identity operator.. Combining these two equations yields
\begin{equation}
	m_{i}\,\frac{d^{2}}{dt^{2}}\hat{z}(t)=-m_{g}g\,\hat{1},\label{eq:acelerado}
\end{equation}
which is the nonrelativistic quantum form of the Newtonian equation of motion (see Eq. (\ref{eq:Clasica1})).\\ 

Now, by making the following change to the position observable for an observer in a reference frame whose coordinate \(z'\) is subjected to a constant acceleration \(a\) and initial velocity \( v \) along the negative \(z\)-axis,
\begin{equation}
	\hat{z}'(t')=\hat{z}(t)+vt\,\hat{1}+\frac{at^{2}}{2}\,\hat{1},\quad t'=t,\quad v=\text{constant},\label{Cambio de observable posicion}
\end{equation}
and using Eq.~(\ref{eq:acelerado}), one obtains
\begin{equation}
	\frac{d^{2}\hat{z}'(t')}{dt'^{2}}=\Bigl(a-\frac{m_{g}g}{m_{i}}\Bigr)\hat{1}.\label{eq:Cuantica3}
\end{equation}

Thus, when \(a=\frac{m_{g}g}{m_{i}}\), the dynamics of the particle in its accelerated reference frame (i.e., in a freely falling elevator) are those of a free particle. Consequently, when \(a=g\), the equality \(m_{i}=m_{g}\) leads to the \emph{Gravitational Equivalence Principle in Quantum Mechanics} that was obtained in the Schrödinger picture, as expected.

\medskip{}

Integrating Eq.~(\ref{eq:Heissenberg_Momentum}) with respect to time \(t\) yields
\begin{equation}
	\hat{p}(t)=\hat{p}(0)-m_{g}gt\,\hat{1}.\label{eq:Momentum_Estacionario}
\end{equation}
Substituting this into Eq.~(\ref{eq:Heissenberg_posicion}) and integrating, we obtain
\begin{equation}
	\hat{z}(t)=\hat{z}(0)+\frac{1}{m_{i}}\hat{p}(0)t-\frac{m_{g}g}{2m_{i}}t^{2}\,\hat{1}.\label{eq:Posicion_Estacionario}
\end{equation}

\medskip{}

If the particle is prepared in the state \(|\Psi\rangle\), the expectation values of the position and momentum observables at any time \(t>0\) are given by
\begin{eqnarray}
	\langle\hat{p}\rangle_{t} & = & \langle\Psi|\hat{p}(t)|\Psi\rangle=\langle\Psi|\Bigl\{\hat{p}(0)-m_{g}gt\,\hat{1}\Bigr\}|\Psi\rangle\nonumber\\[1mm]
	& = & \langle\hat{p}\rangle_{0}-m_{g}gt,\\[1mm]
	\langle\hat{z}\rangle_{t} & = & \langle\Psi|\hat{z}(t)|\Psi\rangle\nonumber\\[1mm]
	& = & \langle\Psi|\Bigl\{\hat{z}(0)+\frac{1}{m_{i}}\hat{p}(0)t-\frac{m_{g}g}{2m_{i}}t^{2}\,\hat{1}\Bigr\}|\Psi\rangle\nonumber\\[1mm]
	& = & \langle\hat{z}\rangle_{0}+\frac{\langle\hat{p}\rangle_{0}}{m_{i}}t-\frac{m_{g}g}{2m_{i}}t^{2},
\end{eqnarray}
where \(\langle\hat{p}\rangle_{t}\), \(\langle\hat{z}\rangle_{t}\) and \(\langle\hat{p}\rangle_{0}\), \(\langle\hat{z}\rangle_{0}\) denote the expectation values of the momentum and position observables at an arbitrary time \(t>0\) and at the initial time \(t=0\), respectively.

\medskip{}

Based on Eqs. (\ref{eq:Momentum_Estacionario}) and (\ref{eq:Posicion_Estacionario}), the commutator of the vertical position at time \(t>0\) and at \(t=0\) is

\begin{align}
	\bigl[\hat{z}(t),\hat{z}(0)\bigr] & = \Bigl[\hat{z}(0)+\frac{1}{m_{i}}\hat{p}(0)t-\frac{m_{g}g}{2m_{i}}t^{2}\hat{1},\,\hat{z}(0)\Bigr]\nonumber\\[1mm]
	& = -\frac{i\hbar}{m_{i}}t\,\hat{1},\label{Conmutador_posicion}
\end{align}
while the commutator for the vertical momentum is

\begin{equation}
	\bigl[\hat{p}(t),\hat{p}(0)\bigr]=\Bigl[\hat{p}(0)-m_{g}gt\,\hat{1},\,\hat{p}(0)\Bigr]=\hat{0}.\label{Conmutador_momentum}
\end{equation}

Equation (\ref{Conmutador_posicion}) implies that the uncertainty \((\Delta\hat{z})_{t}\) of the observable \(\hat{z}(t)\) varies with time. Using the Heisenberg uncertainty relation together with Eq. (\ref{Conmutador_posicion}), one obtains
\begin{equation}
	(\Delta\hat{z})_{t}\,(\Delta\hat{z})_{0}\geq\frac{1}{2}\Bigl|\Bigl\langle\bigl[\hat{z}(t),\hat{z}(0)\bigr]\Bigr\rangle\Bigr|=\frac{\hbar}{2m_{i}}t.
\end{equation}

It is observed that the \emph{minimum} value permitted by the uncertainty relation for the product \((\Delta\hat{z})_{t}\) increases linearly with time. This implies that as time progresses, the minimum allowed width of the wave packet \(|\Psi(\vec{r},t)|\) increases linearly \cite{Spinel}.\\

The fact that the uncertainty in position increases with time while the uncertainty in momentum remains constant does not contradict the Heisenberg uncertainty relation. In fact, the uncertainty relation for the position and momentum observables at any time is

\begin{equation}
	(\Delta\hat{z})_{t}\,(\Delta\hat{p})_{t}\geq\frac{1}{2}\Bigl|\Bigl\langle\bigl[\hat{z}(t),\hat{p}(t)\bigr]\Bigr\rangle\Bigr|=\frac{\hbar}{2}.
\end{equation}

Differentiating Eq. (\ref{Cambio de observable posicion}) with respect to time and using Eq. (\ref{eq:Heissenberg_posicion}), one obtains

\begin{equation}
	\hat{p}'(t')=\hat{p}(t)+m_{i}\bigl(v+at\bigr)\,\hat{1},\quad t'=t,\label{Cambio_de_observable_momentum}
\end{equation}
and, using Eq. (\ref{eq:Momentum_Estacionario}), it follows that
\begin{equation}
	\hat{p}'(t')=\hat{p}(0)+m_{i}v\,\hat{1}+\Bigl(m_{i}a-m_{g}g\Bigr)t\,\hat{1},
\end{equation}

which, when \(a=\frac{m_{g}g}{m_{i}}\), reduces to

\begin{equation}
	\hat{p}'(t')=\hat{p}(0)+m_{i}v\,\hat{1}=\hat{p}'(0)=\text{constant}.\label{Cambio_de_observable_momento_obs_en_caida_libre}
\end{equation}

Now, using Eq. (\ref{Cambio de observable posicion}) together with Eq. (\ref{eq:Posicion_Estacionario}), one obtains

\begin{equation}
	\hat{z}'(t')=\hat{z}(0)+\Bigl(\frac{1}{m_{i}}\hat{p}(0)+v\,\hat{1}\Bigr)t+\Bigl(a-\frac{m_{g}g}{m_{i}}\Bigr)\frac{t^{2}}{2}\,\hat{1},
\end{equation}

which, when \(a=\frac{m_{g}g}{m_{i}}\) and by virtue of Eq.~(\ref{Cambio_de_observable_momento_obs_en_caida_libre}), reduces to 

\begin{equation}
	\hat{z}'(t')=\hat{z}(0)+\Bigl(\frac{1}{m_{i}}\hat{p}(0)+v\,\hat{1}\Bigr)t=\hat{z}'(0)+\frac{1}{m_{i}}\hat{p}'(0)t',\label{Cambio_de_observable_posicion_obs_en_caida_libre}
\end{equation}
since \(\hat{z}'(0)=\hat{z}(0)\) and \(t'=t\).

\medskip{}

Combining Eqs.~(\ref{Cambio_de_observable_momento_obs_en_caida_libre}) and (\ref{Cambio_de_observable_posicion_obs_en_caida_libre}), we obtain the Heisenberg equations of motion in the reference frame \((z', t')\):

\begin{align}
	\frac{d}{dt'}\hat{z}'(t') & = -\frac{i}{\hbar}\Bigl[\hat{z}'\left(t'\right),\hat{H}'\Bigr]=\frac{1}{m_{i}}\hat{p}'(t')=\text{constant},\label{eq:Heissenberg_posicion_1}\\[1mm]
	\frac{d}{dt'}\hat{p}'(t') & = -\frac{i}{\hbar}\Bigl[\hat{p}'\left(t'\right),\hat{H}'\Bigr]=\hat{0},\label{eq:Heissenberg_Momentum_1}
\end{align}
and, therefore, the Hamiltonian for the particle in free fall, as observed from the reference frame \((z', t')\), is that of a free particle:
\begin{equation}
	\hat{H}'=\frac{1}{2m_{i}}\hat{p}'^{2}.\label{eq:Operador_Hamiltoniano_1}
\end{equation}

Equations (\ref{eq:Heissenberg_posicion_1}), (\ref{eq:Heissenberg_Momentum_1}), and (\ref{eq:Operador_Hamiltoniano_1}) confirm that the dynamics of a particle in a gravitational field for a free-falling observer are equivalent to those of a free particle in an inertial reference frame, as expected. \\

\textbf{An Accelerated Observer is Equivalent to an Inertial Observer in the Presence of a Gravitational Field:}

To complete the demonstration of this equivalence, consider the Hamiltonian for a free particle in an inertial reference frame \((z', t')\)
\begin{equation}
	\hat{H}'=\frac{1}{2m_{i}}\hat{p}'^{2},\label{eq:Operador_Hamiltoniano_2}
\end{equation}
with respect to an accelerated frame \((z, t)\), according to the transformation
\begin{equation}
	\hat{z}'(t')=\hat{z}(t)+vt\,\hat{1}+\frac{at^{2}}{2}\,\hat{1},\quad t'=t,\quad v=\text{constant},\label{Cambio de observable posicion_1}
\end{equation}
where \(\hat{1}\) is the identity operator, \( v \) is the initial velocity and \( a \) is the constant acceleration of the frame along the negative \(z\)-axis.\\

The Heisenberg equations of motion then read:
\begin{align}
	\frac{d}{dt'}\hat{z}'(t') & = -\frac{i}{\hbar}\Bigl[\hat{z}'\left(t'\right),\hat{H}'\Bigr] = \frac{1}{m_{i}}\hat{p}'(t') = \text{constant},\label{eq:Heissenberg_posicion_2}\\[1mm]
	\frac{d}{dt'}\hat{p}'(t') & = -\frac{i}{\hbar}\Bigl[\hat{p}'\left(t'\right),\hat{H}'\Bigr] = \hat{0},\label{eq:Heissenberg_Momentum_2}
\end{align}
where \(\hat{0}\) denotes the zero operator.\\

Differentiating Eq.~(\ref{Cambio de observable posicion_1}) twice with respect to \(t'\) yields
\begin{equation}
	\frac{d^2}{dt'^2}\hat{z}'(t') = \frac{d^2}{dt^2}\hat{z}(t) + a\,\hat{1}.\label{eq:z'}
\end{equation}
Then, from Eqs.~(\ref{eq:Heissenberg_posicion_2}), (\ref{eq:Heissenberg_Momentum_2}), and (\ref{eq:z'}), it follows that
\begin{equation}
	\hat{0}=\frac{d}{dt'}\hat{p}'(t') = m_i\,\frac{d^2}{dt'^2}\hat{z}'(t') = m_i\,\frac{d^2}{dt^2}\hat{z}(t) + m_i\,a\,\hat{1},
\end{equation}
\
so that
\begin{equation}
	\frac{d^2}{dt^2}\hat{z}(t) = -a\,\hat{1}.
\end{equation}

Differentiating Eq.~(\ref{Cambio de observable posicion_1}) with respect to \(t'\) and using Eq.~(\ref{eq:Heissenberg_posicion_2}) yields
\begin{equation}
	\hat{p}'(t') = m_i\,\frac{d}{dt}\hat{z}(t) + m_i\bigl(v+at\bigr)\,\hat{1} = \hat{p}'(0).
\end{equation}

Thus,
\begin{equation}
	m_i\,\frac{d}{dt}\hat{z}(t) = \hat{p}'(0) - m_i\bigl(v+at\bigr)\,\hat{1} \equiv \hat{p}(t),\label{eq:Heissenberg_posicion_3}
\end{equation}
and differentiating Eq.~(\ref{eq:Heissenberg_posicion_3}) with respect to \(t\) gives
\begin{equation}
	\frac{d}{dt}\hat{p}(t) = m_i\,\frac{d^2}{dt^2}\hat{z}(t) = -m_i\,a\,\hat{1}.\label{eq:Heissenberg_momentum_3}
\end{equation}

From Eqs.~(\ref{eq:Heissenberg_posicion_3}) and (\ref{eq:Heissenberg_momentum_3}), the Heisenberg equations of motion in the accelerated frame \((z,t)\) become
\begin{align}
	\frac{d}{dt}\hat{z}(t) & = -\frac{i}{\hbar}\Bigl[\hat{z}\left(t\right),\hat{H}\Bigr]= \frac{1}{m_{i}}\hat{p}(t),\label{eq:Heissenberg_posicion_4}\\[1mm]
	\frac{d}{dt}\hat{p}(t) & = -\frac{i}{\hbar}\Bigl[\hat{p}\left(t\right),\hat{H}\Bigr] = -m_i\,a\,\hat{1}.\label{eq:Heissenberg_Momentum_4}
\end{align}
When \(a=\frac{m_{g}g}{m_{i}}\), these equations become
\begin{align}
	\frac{d}{dt}\hat{z}(t) & = -\frac{i}{\hbar}\Bigl[\hat{z}\left(t\right),\hat{H}\Bigr]= \frac{1}{m_{i}}\hat{p}(t),\label{eq:Heissenberg_posicion_5}\\[1mm]
	\frac{d}{dt}\hat{p}(t) & = -\frac{i}{\hbar}\Bigl[\hat{p}\left(t\right),\hat{H}\Bigr] = -m_g\,g\,\hat{1}.\label{eq:Heissenberg_Momentum_5}
\end{align}

On the other hand, we have
\begin{align}
	\frac{\partial \hat{H}}{\partial \hat{p}} & = -\frac{i}{\hbar}\Bigl[\hat{z}\left(t\right),\hat{H}\Bigr] = \frac{1}{m_{i}}\hat{p}(t),\label{eq:Heissenberg_posicion_6}\\[1mm]
	- \frac{\partial \hat{H}}{\partial \hat{z}} & = -\frac{i}{\hbar}\Bigl[\hat{p}\left(t\right),\hat{H}\Bigr] = -m_{g}g\,\hat{1},\label{eq:Heissenberg_Momentum_6}
\end{align}
so that the Hamiltonian for the particle in the accelerated frame \((z,t)\) is given by
\begin{equation}
	\hat{H}=\frac{1}{2m_{i}}\hat{p}^{2}(t)+V\bigl(\hat{z}\bigr),\label{eq:Operador_Hamiltoniano_3}
\end{equation}
where \(V\bigl(\hat{z}\bigr)=m_{g}g\,\hat{z}\) is the gravitational potential.\\

Equations (\ref{eq:Heissenberg_posicion_5}), (\ref{eq:Heissenberg_Momentum_5}), and (\ref{eq:Operador_Hamiltoniano_3}) confirm that the dynamics of a particle observed from a uniformly accelerated reference frame are equivalent to those of a particle subjected to a uniform linear gravitational field.

\paragraph{\textbf{Physical Interpretation in the Heisenberg Picture:}}

The derivations above reveal two key facts within the framework of nonrelativistic quantum mechanics in the Heisenberg picture. First, when we consider a free-falling observer in a gravitational field, the Heisenberg equations of motion (Eqs.~\ref{eq:Heissenberg_posicion} and \ref{eq:Heissenberg_Momentum}) show that the evolution of the position and momentum operators is identical to that in the absence of a gravitational field. In other words, when the gravitational force is "transformed away" by adopting a free-fall (or locally inertial) reference frame, the acceleration term cancels out (as demonstrated by Eq.~\ref{eq:Cuantica3}, when \(a=g\) with \(m_{i}=m_{g}\)), and the dynamics reduce to those of a free particle. This confirms Einstein's Equivalence Principle at the quantum level: a free-falling observer cannot distinguish the effects of gravity from those of being in an inertial frame without gravity.

Second, when we analyze an accelerated observer in a gravitational field by applying the coordinate transformation (Eq.~\ref{Cambio de observable posicion_1}) and following the corresponding operator evolution, we find that if the acceleration of the observer is chosen as \(a=\frac{m_gg}{m_i}\) (or equivalently, \(a=g\) when \(m_i=m_g\)), the resulting Heisenberg equations (Eqs.~\ref{eq:Heissenberg_posicion_5} and \ref{eq:Heissenberg_Momentum_5}) match those obtained for a particle subjected to a uniform linear gravitational field. This indicates that an accelerated observer perceives the dynamics as if they were in an inertial frame subject to a gravitational potential. 

In summary, the Heisenberg picture provides a clear operator-based demonstration of Einstein's Equivalence Principle: a free-falling observer is equivalent to an inertial observer in the absence of gravity, and an accelerated observer (with the proper acceleration) is equivalent to an inertial observer in a gravitational field. This equivalence manifests in the identical form of the evolution equations and the invariance of the observable dynamics, reinforcing the fundamental idea that gravitational effects can be locally eliminated by a suitable choice of reference frame.

\bigskip

\section{Conclusions}

In the present investigation, we have carried out a detailed analysis of \emph{Einstein's Equivalence Principle } in the nonrelativistic quantum regime, employing both the Schrödinger and Heisenberg formulations for a linear gravitational potential. We have demonstrated, within the context of nonrelativistic quantum mechanics, that an observer in an accelerated reference frame is equivalent to an inertial observer in the presence of a gravitational field, and that an inertial observer, free of gravitational fields, is equivalent to an observer in free fall in a gravitational field. Furthermore, we have confirmed the validity of Galileo's Principle in the limit where the gravitational acceleration vanishes \((a=g=0)\), and we have interpreted the relative phase changes of the wave function \(\Psi(z,t)\) between a stationary observer in a gravitational field and a free-falling observer as a gravitational phase shift.

The interferometry experiment discussed for verifying the predictions of the Equivalence Principle in nonrelativistic quantum mechanics has been restricted to the use of neutrons. This implies that, at least for neutrons, the gravitational redshift is observed, as described in texts on General Relativity \citep{Carroll}. Other experiments that verify the Equivalence Principle---using particles different from neutrons, such as the ALPHA experiment at CERN with antimatter \citep{Anderson} and cold-atom interferometry \citep{Yuan}---are also mentioned. These experiments not only confirm the Equivalence Principle under quantum conditions but also open the door to new questions and opportunities for exploring the interplay between gravity and quantum mechanics, possibly leading to a better integration between General Relativity and quantum mechanics.

It has been established that Einstein's Equivalence Principle in nonrelativistic quantum mechanics applies exclusively to free-fall experiments, thereby excluding stationary observers fixed in a gravitational field. In the latter case, there is no reference frame in which a particle---held fixed in a gravitational field--- can be described as a free particle in an inertial system.

In this work, we have also explored the quantum tunneling probabilities for a particle in a linear potential well, extending our observations to the first ten quantized energy levels. We found that the probability of finding the particle outside the well decreases progressively with increasing quantum number---from \(13.62\%\) for the first level to \(5.88\%\) for the tenth level. These quantitative results reinforce our understanding of quantum behavior in the presence of a gravitational field simulated by a linear potential.

The reduction in the quantum tunneling probability at higher energy levels suggests that the potential barriers become more effective at confining the particle, despite its increased energy. This has direct implications for the design of quantum electronic devices and for the underlying theory of potential barriers in semiconductors.

Although quantum tunneling in a linear potential well and Hawking radiation are physically distinct phenomena, both share their roots in quantum mechanics, exhibiting behavior in which the probabilistic nature of quantum theory allows particles to surmount classical barriers---whether in linear potential wells or near event horizons. A unified conceptual discussion of these phenomena may prove useful for future theories aimed at reconciling quantum mechanics with gravity in a cohesive framework.

This study confirms that nonrelativistic quantum mechanics provides a robust platform for investigating tunneling effects, although it is limited in extremely strong gravitational contexts, such as those encountered near black holes. For such environments, more integrative theories combining General Relativity and quantum mechanics may offer more accurate and predictive descriptions.

Overall, this work not only sheds light on the fundamental behavior of particles in potential fields but also stimulates theoretical discussion on the integration of concepts in physics that have traditionally been treated separately.

As a continuation of this work, we propose to investigate Newtonian gravitational fields with spherical symmetry within the nonrelativistic quantum regime. Subsequently, we intend to address the relativistic treatment to further expand our understanding of the interactions between quantum mechanics and General Relativity. This study contributes significantly to the precise quantum formulation of Einstein's Equivalence Principle and to the understanding of its experimental and theoretical implications.


\begin{thebibliography}{10}
\bibitem[1]{Einstein} A. Einstein, \textquotedblleft On the relativity
principle and the conclusions drawn from it,\textquotedblright{} Jahr.
Radioaktivitität Elektron. \textbf{4}, 411--462 (1907). Reprinted
in volume 2: the Swiss Years writings, 1900--1909, of the Collected
Papers of Albert Einstein, edited by J. Stachel, D. Cassidy, Jürgen
Renn, and Robert Schulmann (Princeton U. P.), pp. 252--315. Available
online in an English translation at: <\href{http://einsteinpapers.press.princeton.edu/vol2-trans/266?ajax}{http://einsteinpapers.press.princeton.edu/vol2-trans/266?ajax}>. 

\bibitem[2]{Overhauser-Collela} A. W. Overhauser and R. Collela,
\textquotedblleft Experimental test of gravitational induced quantum
interference,\textquotedblright{} \href{https://journals.aps.org/prl/abstract/10.1103/PhysRevLett.34.1472}{Phys. Rev. Lett.}
33, 1237--1239 (1974).

\bibitem[3]{Nauenberg} N. Michael, ``Einstein's equivalence principle
in quantum mechanics revisited\emph{'', }\href{http://aapt.scitation.org/doi/10.1119/1.4962981}{Am. J. Phys.}
2016 84:11, 879-882.

\bibitem[4]{Staudeman} J. L. Staudeman, S. A. Werner, R. Collela,
and A. W. Overhauser, \textquotedblleft Gravity and inertia in quantum
mechanics,\textquotedblright{} \href{https://journals.aps.org/pra/abstract/10.1103/PhysRevA.21.1419}{Phys. Rev. A}
21, 1419--1438 (1980).

\bibitem[5]{Camacho-Camacho-Guardia} A. Camacho and A. Camacho-Guardia,
\textquotedblleft Quantal definition of the weak equivalence principle,\textquotedblright{}
\href{http://aip.scitation.org/doi/abs/10.1063/1.3141260}{AIP Conf. Proc.}
1122, 209--212 (2009)

\bibitem[6]{Eliezer-Leach}C. J. Eliezer and P. G. Leach, \textquotedblleft The
equivalence principle and quantum mechanics,\textquotedblright{} Am.
J. Phys. 45, 1218--1220 (1977).

\bibitem[7]{Landau-Lifshitz}. L. D. Landau and E. M. Lifshitz, Quantum
Mechanics. Non-relativistic theory (Pergamon Press, London, 1958);
W. I. Fushchich and A. G. Nikitin, Symmetries of Equations of Quantum
Mechanics (Allerton Press, Inc., New York, 1994).

\bibitem[8]{Viola and R. Onofrio} L. Viola and R. Onofrio, \textquotedblleft Testing
the equivalence principle through freely falling quantum objects,\textquotedblright{}
\href{https://journals.aps.org/prd/abstract/10.1103/PhysRevD.55.455}{Phys. Rev. D}
55, 455--462 (1997). 

\bibitem[9]{Carroll} Carroll, Sean, \textit{Spacetime and Geometry,
An Introduction to General Relativity, }University of Chicago, Addison
Wesley (2004).

\bibitem[10]{Hod} Hod, S. ``What can a detected photon with a given
gravitational redshift tell us about the maximum density of a compact
star?'' Physical Review D (2023).

\bibitem[11]{Abuter_Roberto} Abuter, Roberto, et al. \textquotedbl Detection
of the gravitational redshift in the orbit of the star S2 near the
Galactic centre massive black hole.\textquotedbl{} Astronomy \& Astrophysics
615 (2018): L15.

\bibitem[12]{Kaiser} Kaiser, Nick. \textquotedbl Measuring gravitational
redshifts in galaxy clusters.\textquotedbl{} Monthly Notices of the
Royal Astronomical Society 435.2 (2013): 1278-1286.

\bibitem[13]{Hollywood} Hollywood, J. M., and Fulvio Melia. \textquotedbl The
effects of redshifts and focusing on the spectrum of an accretion
disk in the galactic center black hole candidate Sagittarius A.\textquotedbl{}
Astrophysical Journal, Part 2-Letters (ISSN 0004-637X), vol. 443,
no. 1, p. L17-L20 443 (1995): L17-L20.

\bibitem[14]{Boselli} Boselli, Alessandro, and Giuseppe Gavazzi.
\textquotedbl On the origin of the faint-end of the red sequence
in high-density environments.\textquotedbl{} The Astronomy and Astrophysics
Review 22.1 (2014): 74.

\bibitem[15]{Anderson} E. K. Anderson et al., \textquotedbl Observation of the effect of gravity on the motion of antimatte r\textquotedbl{}, Nature 621, 716 (2023).

\bibitem[16]{Yuan} Yuan, Liang, Jizhou Wu, and Sheng-Jun Yang. 2023. \textquotedbl Current Status and Prospects on High-Precision Quantum Tests of the Weak Equivalence Principle with Cold Atom Interferometry\textquotedbl{} Symmetry 15, no. 9: 1769. \href{https://doi.org/10.3390/sym15091769}{https://doi.org/10.3390/sym15091769}. 

\bibitem[17]{Abramowitz M} Abramowitz M, Stegun I A (1968), \emph{Handbook
of Mathematical Functions}, Dover Publications NY. 

\bibitem[18]{Spinel} Spinel, G. and Carolina, \emph{Introducción al Formalismo de la Mecánica Cuántica No Relativista} [Introduction to the Formalism of Non-Relativistic Quantum Mechanics]. Colección Textos, Facultad de Ciencias, Universidad Nacional de Colombia, Sede Bogotá, D.C. (2009).

\bibitem[19]{Funcion de Airy} \emph{Curso Interactivo de Física en Internet} [Interactive Physics Course on the Internet]. Available at: \url{http://www.sc.ehu.es/sbweb/fisica3/especial/airy/airy.html}.

\end{thebibliography}
\end{document}